\documentclass[12pt]{iopart}
\expandafter\let\csname equation*\endcsname\relax
\expandafter\let\csname endequation*\endcsname\relax
\usepackage{amsmath}
\usepackage{amsfonts}
\usepackage{amssymb}
\usepackage[scanall]{psfrag}
\usepackage{psfrag}
\usepackage{graphicx,caption,float}
\usepackage{color}
\usepackage{tabularx}
\usepackage{blindtext}

\begin{document}
\newcommand{\of}[1]{\left( #1 \right)}
\newcommand{\sqof}[1]{\left[ #1 \right]}
\newcommand{\abs}[1]{\left| #1 \right|}
\newcommand{\avg}[1]{\left< #1 \right>}
\newcommand{\cuof}[1]{\left \{ #1 \right \} }
\newcommand{\bra}[1]{\left < #1 \right | }
\newcommand{\ket}[1]{\left | #1 \right > }
\newcommand{\pil}{\frac{\pi}{L}}
\newcommand{\bx}{\mathbf{x}}
\newcommand{\by}{\mathbf{y}}
\newcommand{\bk}{\mathbf{k}}
\newcommand{\bp}{\mathbf{p}}
\newcommand{\bl}{\mathbf{l}}
\newcommand{\bq}{\mathbf{q}}
\newcommand{\psibar}{\overline{\psi}}
\newcommand{\svec}{\overrightarrow{\sigma}}
\newcommand{\dvec}{\overrightarrow{\partial}}
\newcommand{\bA}{\mathbf{A}}
\newcommand{\bdelta}{\mathbf{\delta}}
\newcommand{\bK}{\mathbf{K}}
\newcommand{\bQ}{\mathbf{Q}}
\newcommand{\bG}{\mathbf{G}}
\newcommand{\bw}{\mathbf{w}}
\newcommand{\bL}{\mathbf{L}}
\newcommand{\ohat}{\widehat{O}}
\newcommand{\up}{\uparrow}
\newcommand{\down}{\downarrow}
\newcommand{\MM}{\mathcal{M}}
\newcommand{\tX}{\tilde{X}}
\newcommand{\tY}{\tilde{Y}}
\newcommand{\tZ}{\tilde{Z}}
\newcommand{\tOm}{\tilde{\Omega}}
\newcommand{\barA}{\bar{\alpha}}




\title{Embedding quantum optimization problems using AC driven quantum ferromagnets}

\author{Gianni Mossi$^{1,2}$, Vadim Oganesyan$^{3,4}$, Eliot Kapit$^5$}
\address{$^1$ KBR, Inc., 601 Jefferson St., Houston, TX 77002, USA.}
\address{$^2$ Quantum Artificial Intelligence Lab. (QuAIL), NASA Ames Research Center, Moffett Field, CA 94035, USA.}
\address{$^3$ Department of Physics and Astronomy, College of Staten Island, CUNY, Staten Island, NY 10314, USA}
\address{$^4$ Physics program and Initiative for the Theoretical Sciences, The Graduate Center, CUNY, New York, NY 10016, USA}
\address{$^5$ Department of Physics and Quantum Engineering Program, Colorado School of Mines, Golden, CO 80401}

\begin{abstract}
Analog quantum optimization methods, such as quantum annealing, are promising and at least partially noise tolerant ways to solve hard optimization and sampling problems with quantum hardware. However, they have thus far failed to demonstrate broadly applicable quantum speedups, and an important contributing factor to this is slowdowns from embedding, the process of mapping logical variables to long chains of physical qubits, enabling arbitrary connectivity on the short-ranged 2d hardware grid. Beyond the spatial overhead in qubit count, embedding can lead to severe time overhead, arising from processes where individual chains ``freeze" into ferromagnetic states at different times during evolution, and once frozen the tunneling rate of this single logical variable decays exponentially in chain length. We show that this effect can be substantially mitigated by local AC variation of the qubit parameters as in the RFQA protocol (Kapit and Oganesyan, Quant. Sci. Tech. \textbf{6}, 025013 (2021)), through a mechanism we call Symphonic Tunneling. We provide general arguments and substantial numerical evidence to show that AC-driven multi-qubit tunneling is dramatically faster than its DC counterpart, and since ST is not a 1d-specific mechanism, this enhancement should extend to clusters of coupled chains as well. And unlike a uniform transverse field, in higher dimensions this method cannot be efficiently simulated classically. We explore schemes to synchronize the AC tones within chains to further improve performance. Implemented at scale, these methods could significantly improve the prospects for achieving quantum scaling advantages in near-term hardware. 
\end{abstract}

\maketitle

\section{Introduction}

While enormous progress has been made in gate model quantum computing in recent years, quantum advantage for practical, real-world problems with such devices has yet to be demonstrated. This is largely due to the extreme precision requirements and noise sensitivity inherent to the gate model. In contrast, \emph{analog} quantum optimizers, including quantum annealers (QA) \cite{finnila1994quantum,kadowakinishimori1998,farhigoldstone2000,farhi2002quantum,das2008colloquium,johnson2011quantum,boixo2014evidence,farhi2014quantum,albashlidar2017,albash2018demonstration} and more recent Rydberg atom systems \cite{scholl2021quantum, scholl2022microwave, ebadi2022quantum}, are much more resilient can routinely solve problems with hundreds or even thousands of variables. While not computationally universal, a huge array of NP-hard and NP-complete problems can be solved using these systems, with the likely mechanism of quantum advantage lying in collective quantum tunneling as a way to escape local minima once the system becomes glassy \cite{perdomo2011study,king2018observation,chancellor2017modernizing,marshall2019power}. Recent experimental results \cite{albash2018demonstration,ebadi2022quantum,king2023quantum} on large problems in both neutral atom and flux qubit analog devices suggest quantum scaling advantage over classical heuristics, for problems which are short ranged and map natively to the underlying hardware graph. However, demonstration of \emph{broadly applicable} quantum scaling advantage has remained elusive, particularly for problems defined on abstract graphs with long ranged connectivity. 

To attack these problems, encodings such as minor embedding are required \cite{choi2008minor,choi2011minor,konz2021embedding}, where single logical variables are mapped to long ferromagnetically coupled chains (similar, antiferromagnetic schemes exist for Rydberg arrays \cite{kim2021rydberg,nguyen2023quantum}). These methods have an inescapable quadratic overhead in qubit count, but more insidiously, they can introduce severe time penalties as well \cite{konz2021embedding,marshall2017thermalization,hamerly2019experimental,kowalsky20213}. The essential mechanism for this in large problems is that individual chains will ``freeze" into ferromagnetic states at different times during evolution (note of course that the whole system will eventually freeze as the transverse field driving quantum dynamics is reduced to zero). In the process, transverse field corrections and other effects can tip them into the ``wrong" state, and once frozen the tunneling rate of these single variables decays exponentially in chain length. This leads to an ugly tradeoff in the strength of the intra-chain coupling $J_c$. If it is too weak, the system becomes more noise sensitive and can have (logically meaningless) broken chains in its ground state. Larger $J_c$ values however, tend to freeze earlier and have even worse scaling for multiqubit tunneling. In the worst cases, the time to solution can scale exponentially in the total system size, e.g. $\exp \of{ N^2}$ for $N$ logical variables \cite{hamerly2019experimental}. And unlike the spatial overhead, this problem cannot be solved by simply making the system larger; a more fundamental change to the quantum optimization protocol is required.

\begin{figure}[t]
	\centering
	
	\includegraphics[width=0.85\textwidth]{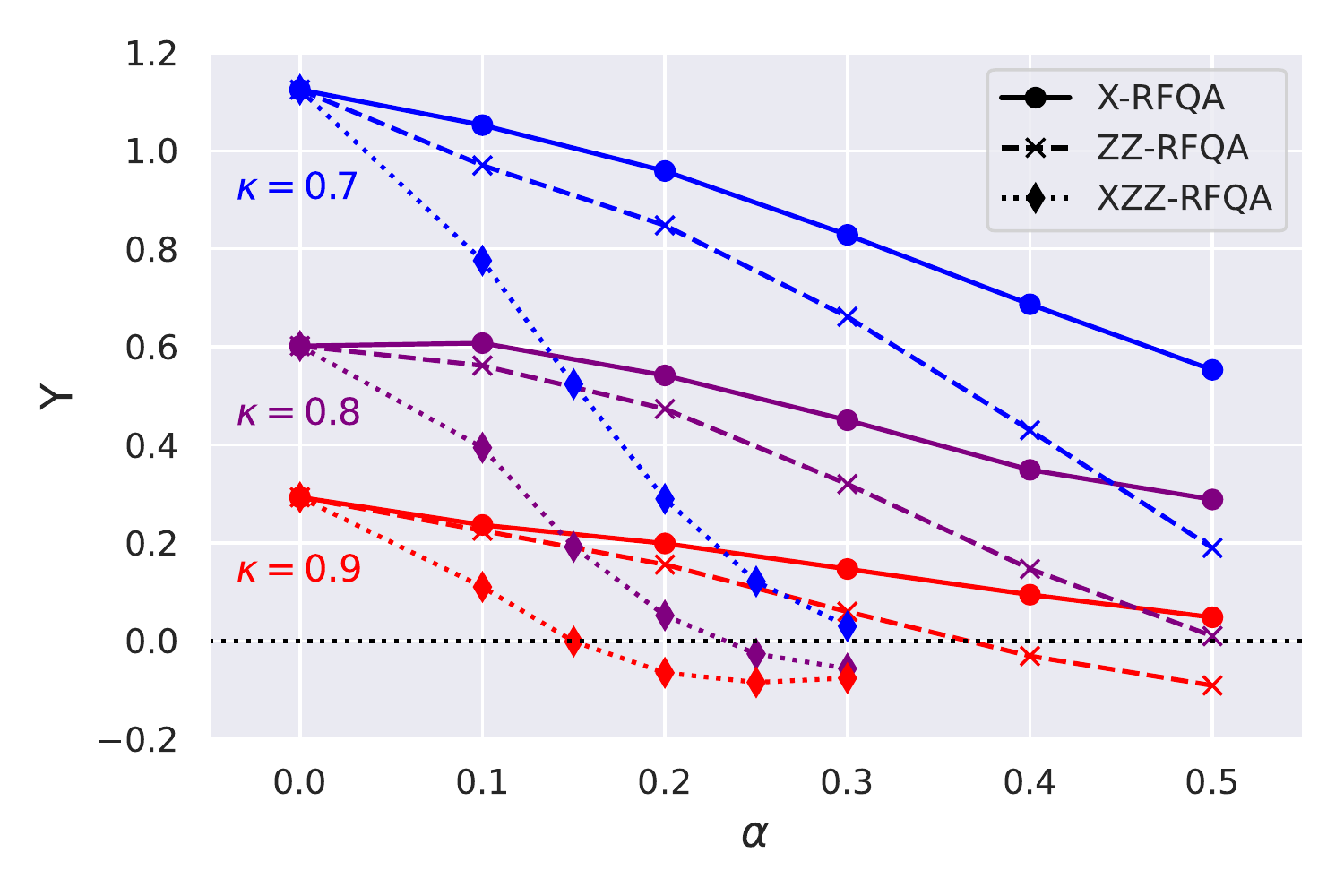}
	
	\caption{Difficulty exponent $\Upsilon$
 obtained from the rate of complete magnetization reversal (see Eq. \ref{eq:gamma_and_upsilon_relation} for definition and Sec. \ref{sec:results} for further details) as a function of oscillations' amplitude $\alpha$ for three different driving protocols defined in Sec. \ref{sec:details}, for three different values of transverse field $\kappa$. A vanishing or negative $\Upsilon$ implies that an exponential fit is no longer good (at least for the ranges simulated), indicating a potential crossover to polynomial scaling.} 
 
	\label{fig:tts_scaling_exp}
\end{figure}

We have demonstrated previously the use of randomised dynamical protocols to ameliorate exponential scaling, both analytically and numerically in simple 0-dimensional problems of $N$ coupled qubits, where level structure exhibits an avoided level crossing with a gap to other excited states. The Random Field (or Radio Frequency) Quantum Annealing (RFQA) protocol\cite{kapit2021noise} utilizes this separation of scales to accelerate the mixing across the level crossing with no appreciable heating, thus reducing the ``difficulty exponent"  
\begin{equation}
\label{eq:gamma_and_upsilon_relation}
    \Upsilon\equiv \frac{1}{N}\log_{2} \Big( \frac{1}{\Gamma (N)} \Big),
\end{equation}
as extracted from the tunnelling rate $\Gamma(N)$ (more on this in the next Section \ref{sec:IsingModelEtc}), which is constrained in those structureless models by the optimal adiabatic value, e.g. $\Upsilon \geq \frac{1}{2}$ for the Grover problem \cite{zalka1999}. While this optimal value is exponentially fragile to control precision, RFQA is not.  By contrast, quantum tunnelling of logical qubits based on the two groundstates of the Ising chain of length $L$ is governed by a highly structured Hamiltonian and we therefore expect to be able to accelerate the dynamics more efficiently.  To this end we explored a number of \emph{structured} protocols inspired both by basic understanding of correlated dynamics of small spin clusters and practical considerations of hardware implementation. Remarkably, we find significant reductions $\Upsilon$, seemingly to arbitrarily small values (without appreciable heating or melting of ferromagnetic order -- see Section \ref{sec:results}), as we vary drive strength and other detials, see Fig. \ref{fig:tts_scaling_exp}.  We refer to the general scheme of structured/optimized multifrequency protocols as Symphonic Tunnelling.

To demonstrate the potential of RFQA/ST to mitigate the time overhead of embedding, we report the results of an extensive series of numerical simulations, focusing on the tunneling of a single one-dimensional ring in the ferromagnetic (frozen) phase, where the base DC tunneling rate decays exponentially in length $L$. We focus on single chains because, relying on AC dynamics, RFQA cannot be simulated with quantum Monte Carlo \cite{isakov2016understanding,andriyash2017can,jiang2017scaling,jiang2017path,king2019scaling}, and thus embedded \emph{problems} are firmly out of reach of classical simulation techniques at any meaningful scales. Likewise, while we expect RFQA methods to accelerate problem solving for native problems, benchmarking that in simulation is problematic because small instances (e.g. $N \leq 20$) typically do not show the expected large-scale exponential difficulty scaling and instances numerically mined for very small gaps are generally fragile to perturbations \cite{crosson2014different}. The exponential scaling of collective tunneling in 1d chains, in contrast, is well-controlled and obvious at small $L$.

Further, the key slowdown mechanism of freezing can be captured in a single chain, and we average over random detuning as a proxy for the disordered environment--and inability to find precise resonances--of real analog problems. And since ST is fundamentally \emph{not} a 1d-specific mechanism, it will generalize to clusters of coupled chains in larger systems. Finally, 1d chains are amenable to matrix product simulation techniques (in this case, time-evolving block decimation \cite{PhysRevLett.93.040502,PhysRevLett.91.147902,PAECKEL2019167998,10.21468/SciPostPhysCodeb.4,10.21468/SciPostPhysCodeb.4-r0.3}), which allow us to extend our simulations out to much larger system sizes than are possible in full-wavefunction evolutions. We demonstrate parameter regimes where the chain remains frozen--in that ferromagnetic correlations do not vanish at large separation--but collective tunneling is accelerated to the point that polynomial and exponential scaling cannot be distinguished from the data.

The rest of this paper is organized as follows. In the next section, we introduce the basic 1d collective tunneling problem and the AC protocols we will use to attack it. Following that, we present details of our numerical simulation methods, including full wavefunction simulations and matrix product methods. We then present numerical results for a range of system parameters and protocols. While we do not have analytical arguments to predict performance in arbitrarily large systems, the robustness of our observed speedups (which persist out to the largest system sizes we can simulate) strongly suggest significant performance improvements at the application scale. We finally detail some of the considerations for a superconducting hardware implementation, and offer concluding remarks.

\section{Ising Model of reverse annealing -- quasi-static DC protocol}
\label{sec:IsingModelEtc}
We consider quantum dynamics of a single ferromagnetic ring, with total Hamiltonian
\begin{eqnarray} \label{eq:Hamiltonian}
H \of{t} = - \sum_j \sqof{\kappa_j \of{t} X_j + h_j \of{t} Z_j + J_j \of{t} Z_j Z_{j+1} },
\end{eqnarray}
with time dependent couplings chosen to accelerate the mixing of two ferromagnetic ground states without any appreciable excitation of excited states. For ease of comparison we will benchmark all protocols using the standard "reverse annealing" quench, whereby we initialize the system in a definite all-down classical groundstate with no transverse field, turn the transverse field on and off slowly, and examine the probability of complete magnetization reversal
\begin{equation}
P(t_f) \equiv \lvert \langle \uparrow \cdots \uparrow \rvert U(t) \lvert \downarrow \cdots \downarrow \rangle \rvert^2.
    \end{equation}
Importantly, we take great precautions to avoid exciting the problem out of the low energy ferromagnetic doublet, i.e. 
\begin{equation}\label{eq:heating}
P_{heat}\equiv 1-\lvert \langle \uparrow \cdots \uparrow \rvert U(t) \lvert \downarrow \cdots \downarrow \rangle \rvert^2-\lvert \langle \downarrow \cdots \downarrow \rvert U(t) \lvert \downarrow \cdots \downarrow \rangle \rvert^2 \ll P(t_f).
\end{equation}
We do not attempt to measure heating during evolution, but since no cooling mechanism is present to remove stray excitations, measuring it at the end is sufficient.

\begin{figure}
\centering
\includegraphics[width=.85\textwidth]{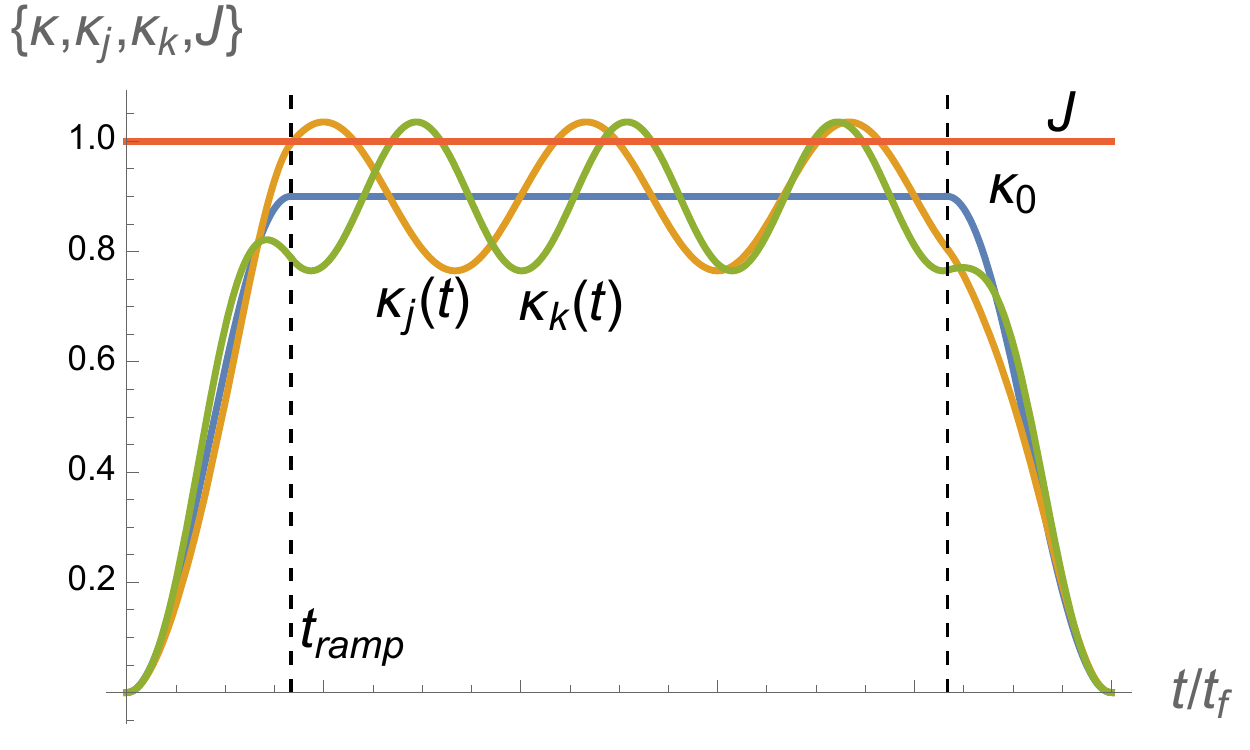}
\caption{The reverse annealing schedule explored in this work. The ferromagnetic coupling $J$ is held fixed (red line), and the uniform transverse field is ramped up and down to induce collective tunneling (blue curve). In the RFQA cases, these parameters are locally modulated with independent oscillating frequencies; the modulation of two (out of $L$) transverse fields $\kappa_j \of{t}$ and $\kappa_k \of{t}$ is shown in gold and green. Note in other protocols individual coupling strengths $J_j $ can be oscillated, and/or these coupling strengths can be modulated by the transverse field strengths of the coupled qubits (as in real quantum annealers).}\label{fig:RA}
\end{figure}

In what we follows we focus on the statistically significant average probably of success, whereby several instances of the problem are considered with resultant $P(t_f)$ averaged over the ensemble of static and dynamic variations of potentials. We average $P \of{t_f}$ over fixed detuning $h$ (corresponding to a uniform $Z$ bias $h/2L$) drawn from the uniform range $\cuof{-2/\sqrt{L},2/\sqrt{L}}$ (this scaling choice is justified below), for both the DC and RFQA cases. We calibrate the analysis of complicated protocols defined below against the simplest uniform quasi-state sweep characterized by the maximum value of transverse field $\kappa_0$ -- see Fig. \ref{fig:RA} and, importantly, the value of detuning $W\equiv 2 \sum_i h_i$.  In this simple case we can accurately estimate (see App. A of \cite{kapit2021noise})
\begin{equation}
    P(t_f)\approx \int d W \mathcal{P}(W) \frac{\Omega_0^2}{W}\sin^2 (W t_f)\to \pi \frac{\Omega_0^2}{W}t_f\equiv \Gamma_0 t_f,
\end{equation}
which ignores ramp time $t_{ramp}$ (in practice, we fix $t_f=6(2\pi N)=6 t_{ramp}$) and assumes are relatively smooth distribution of detuning $\mathcal{P}(W)$ and sufficiently long $t_f$ to sample it efficiently. The most important part of this expression is the many-body matrix element connecting the two magnetization states which for a uniform field case is well-captured by the scaling form
\begin{eqnarray}\label{Om0}
\Omega_0 \of{L} \propto \frac{\kappa_0}{L^{\kappa_0/J}} \of{\frac{\kappa_0}{J}}^{L-1}, \; \; \Gamma_0 \propto \frac{\Omega_0^2}{W}.
\end{eqnarray}
This matrix element figures prominently in the standard textbook formulation of symmetry restoration via path integral instantons, but can also be obtained by direct high order perturbation theory; at large $L$ the polynomial prefactor is largely irrelevant but included for completeness. Since $\Omega_0(L)$ is normally exponentially small in $L$, the total number of spins flipping, we can use the difficulty exponent $\Upsilon = (1/L) \log_{2}(1/\Gamma (L))$ to help isolate the scaling advantage of various protocols. Thus, the uniform field protocol discussed thus far corresponds to $\Upsilon_0= 2 \log_{2} J/\kappa_0 $, which only vanishes at the quantum critical point, where the order parameter vanishes, and with it the ability to use the chains for embedding logical qubits! 

Before turning to the more powerful protocols that appear to deliver dramatic reduction in $\Upsilon$ without destroying the logical qubits in the process we close this section by defining two useful quantitative tools. One is the so called "time-to-solution" (TTS) that is a nice proxy\cite{PhysRevA.92.042325} for $\Gamma$
\begin{eqnarray}\label{defTTS}
   \mathrm{TTS} \equiv t_f \frac{\ln(1-0.99)}{\ln(1-\langle P(t_f) \rangle)} \sim \frac{t_f}{\langle P(t_f) \rangle}\to 1/\Gamma. 
\end{eqnarray}
And finally, we also track the time-averaged order parameter correlation function \emph{during} the waiting stage at largest separation available 
\begin{equation}\label{eq:correlation_function}
    C_{ZZ} = \frac{1}{t_{wait}} \int_{t_{ramp}}^{t_f-t_{ramp}}  \overline{\langle Z_0(t) Z_{L/2}(t)\rangle}  \,\mathrm{d}t .
\end{equation}
Tracking this quantity allows us to confirm that we do not leave the ferromagnetic phase in AC evolution, as discussed below in the results section.


\section{Dynamic protocols and other simulation details}
\label{sec:details}
This section details three novel modifications to the standard quasi-static approach described above. All these approaches include modulations to the three types of couplings already present in the Hamiltonian that are both weak and slow so as to avoid exciting (heating) the system while dramatically increasing the rate of mixing of the two ferromagnetic groundstates. The seemingly inevitable success of this program was demonstrated by two of us both analytically and numerically in the context of featuresless "oracle" problems\cite{kapit2021noise} where the choice of driving terms was randomized and averaged over, which each qubit assigned its own driver. By contrast, the protocols we employ in this work are much less random, with total number of independent frequencies scaling as $\sim \sqrt{L}$, i.e. with many qubits sharing the same driver. This modification appears to dramatically improve the performance of Symphonic Tunnelling, while at the same time precluding a straightforward extension of prior analytic results for the fully random ensemble. Because of its pivotal conceptual importance (certainly in our thinking) and clarity, we first review the basic mechanism of the fully random protocol, RFQA, before turning to the correlated variant employed here in the following paragraph.
Consider an exponentially avoided level crossing or a first order phase transition where $K$ spins must flip. Now imagine to every spin we add an oscillating term $O_j \sin 2\pi f_j t$, where $O_j$ is a local operator. At first order, when the energy difference of the two states approaches any $\pm f_j$, the AC perturbation can resonantly mix the states, with a mean Rabi frequency $c \Omega_0$, for some $O \of{1}$ constant $c$. This accelerates the off-resonant tunneling rate by a factor of $K$, but amounts to a dramatic underestimate, since there are also $4 \binom{K}{2}$ two-frequency processes at second order, $8 \binom{K}{3}$ three-frequency terms at third order, and so on; all are generically nonzero. If we assume that the \emph{average} $m$th order process scales as $\Lambda^m \Omega_0$ (for $\Lambda < 1$, e.g. exponential decay in $m$), then a simple incoherent sum of all contributing processes produces a total energy-averaged tunneling rate
\begin{eqnarray}
\Gamma_T \propto  \frac{\Omega_0^2}{W} \sum_{m=0}^{K} 2^m \Lambda^{2m} \binom{K}{m} \propto \frac{\Omega_0^2}{W} \of{1+ 2 \Lambda^2}^K.
\end{eqnarray}
This can favorably shift the scaling exponent of the problem, and of particular interest to us is the case where $\of{1+ 2 \Lambda^2}^K$ grows faster than $\Omega_0^2$ decays and this simple prediction breaks down. For structureless problems examined previously\cite{kapit2021noise} adiabatic scaling acts as a speedlimit, hence breakdown of perturbation theory is unphysical. However, typical problems of interest are structured and one might expect physical consequence of such a breakdown, e.g. a breakdown of exponential scaling of mixing in favor of a faster process. We \emph{numerically} demonstrate such crossovers for multiple protocols below, where exponential and polynomial scaling cannot be distinguished from the data, though we emphasize that these are numerical results and we do not have any analytical guarantees this scaling would persist as $L\to \infty$.  However, the existence of this dynamical regime for $L\approx 40$ is of keen experimental and practical interest to ongoing efforts to engineer quantum platforms.

We now compare the standard uniform-transverse-field reverse annealing protocol (the ``DC protocol'') with three different implementations of the RFQA protocol. These are differentiated by (i) the operators that are being oscillated, and (ii) the distribution of $L$ random frequencies $\{f_j\}$ and initial phases $\{\phi_j\}$ that parametrize these oscillations. We define a parameter $\alpha$, common to all protocols, that fixes the amplitude of the oscillations. Note that this parameter determines the \emph{relative} magnitude of the oscillations and it multiplies the base parameter (e.g. coupling $J_j$ or transverse field $\kappa_j$) in the full time-dependent $H \of{t}$. In all cases the longitudinal bias field $h_j \of{t} = -h$ is kept fixed throughout evolution. The three RFQA protocols we simulate here are:
\begin{itemize}

\item X-RFQA with randomly-paired oscillations: transverse field strengths are locally modulated, so and $J_j(t)=1$ while $\kappa_j(t) = \kappa(t)(1 + \alpha \sin(f_j t + \phi_j))$ where $\kappa(t)$ is the ramp schedule described in the DC protocol. $O(\sqrt{L})$ random frequencies and phases are generated independently and these are randomly assigned to the spins in the chain (each tone is thus repeated $O \of{\sqrt{L}}$ times). This protocol could be implemented in neutral atoms by locally modulating laser intensities.
\item ZZ-RFQA with randomly-paired oscillations: local ferromagnetic coupling strengths are modulated. Here $\kappa_j(t) = \kappa(t)$ is the schedule described in the DC protocol and interactions are oscillated like $J_j(t) = 1 + \alpha \sin(f_j t + \phi_j)$. We sample the random frequencies and phases in the same way as the X-RFQA case. This protocol can be implemented in quantum annealers through local AC flux control of the coupling terms.
\item XZZ-RFQA with randomly-paired oscillations: the transverse fields follow the same schedule $\kappa_j(t) = \kappa(t)(1 + \alpha \sin(f_j t + \phi_j))$ of the X-RFQA protocol, while ferromagnetic interactions are modulated based on the transverse field strengths of the coupled qubits, as $J_j(t)\equiv (1 - \alpha \sin(f_j t + \phi_j))(1 - \alpha \sin(f_{j+1} t + \phi_{j+1})) $ This reflects real flux qubit physics \cite{johnson2011quantum}, where the stronger the transverse field is, the lower the susceptibility to longitudinal flux, so increasing a local transverse field decreases couplings to that qubit. This protocol could be implemented in flux qubit hardware through time-dependent local flux control of transverse field strengths.
\end{itemize}

Notice that in all cases we repeat each tone $O \of{\sqrt{L}}$ times at random spatial locations--$\sqrt{L}/3$, to be precise (when this quantity is not an integer, each tone is repeated either ${\rm floor} \of{\sqrt{L}/3}$ or ${\rm ceiling} \of{\sqrt{L}/3}$ with appropriate probability, and the process halts when all $L$ sites have frequency/phase pairs assigned). We choose this pairing structure for a few reasons. First, for all three protocols, synchronizing tones causes the corresponding AC terms to interfere constructively, increasing the boost to collective tunneling. However, if the same tone is repeated too many times, it can cause large mean value swings in $\kappa$ and/or $J$, leading to phase transitions into the paramagnetic state. This shows up as more significant heating and the vanishing of the ferromagnetic order parameter (defined above), which are both very undesirable in the context of embedded chains for problem solving. We found $O \of{\sqrt{L}}$ pairing to be a kind of ``sweet spot" between reaping the benefits of synchronization without significant heating or phase transitions out of the ground state doublet (and further, relevant for experimental implementation, smaller modulation amplitudes $\alpha$ are required to obtain the same change in $\Upsilon$ with tone repetition). Finally, in a real implementation of $N$ qubits in a 2d lattice, a requirement of only $O \of{\sqrt{N}}$ unique frequencies significantly reduces the signal generation and control complexity.

Each random frequency is drawn from the range $f_j \in \cuof{1/2L,1/L}$ with random phase $\phi_j$. This choice significantly mitigates heating from the AC drives. The $1/\sqrt{L}$ scaling of the detuning range ensures that an $O \of{1}$ fraction of the RFQA trials are driving tunneling between states separated by any energy difference less than $O \of{\sqrt{L} \avg{f_j}}$; as discussed extensively in \cite{kapit2021noise}, multiqubit transitions are only accelerated by RFQA if the energy difference between the competing states falls in the window in which the frequency combinations are dense, consistent with the core mechanism being an exponential proliferation of weak resonances. This slowly decaying ``resonance" condition is in stark contrast to DC protocols (the ``population transfer" or ``reverse annealing" simulated here) where given an exponentially decaying minimum gap $\Omega_0$, $P \of{t_f}$ can only reach appreciable values if the detuning $h$ is $O \of{\Omega_0}$ or less. Finally, to regularize behavior at small $L$--and consequently provide a larger range of reliable data to fit--we also perform our simulations with a single transverse field, at site 0, weakened by a constant prefactor. This reduces the degeneracy splitting $\Omega_0$ by a prefactor without changing the scaling with $L$; absent this step, at larger $\kappa$ values and small $L$, tunneling occurs too quickly to allow us to employ long enough runtimes to use small frequencies and thus, mitigate heating from the AC drives. For fair comparisons the same single field weakening is used in all simulations.

\section{Results}
\label{sec:results}
\subsection{Acceleration of Tunnelling}
\begin{figure}
	\centering
	
	\includegraphics[width=0.85\textwidth]{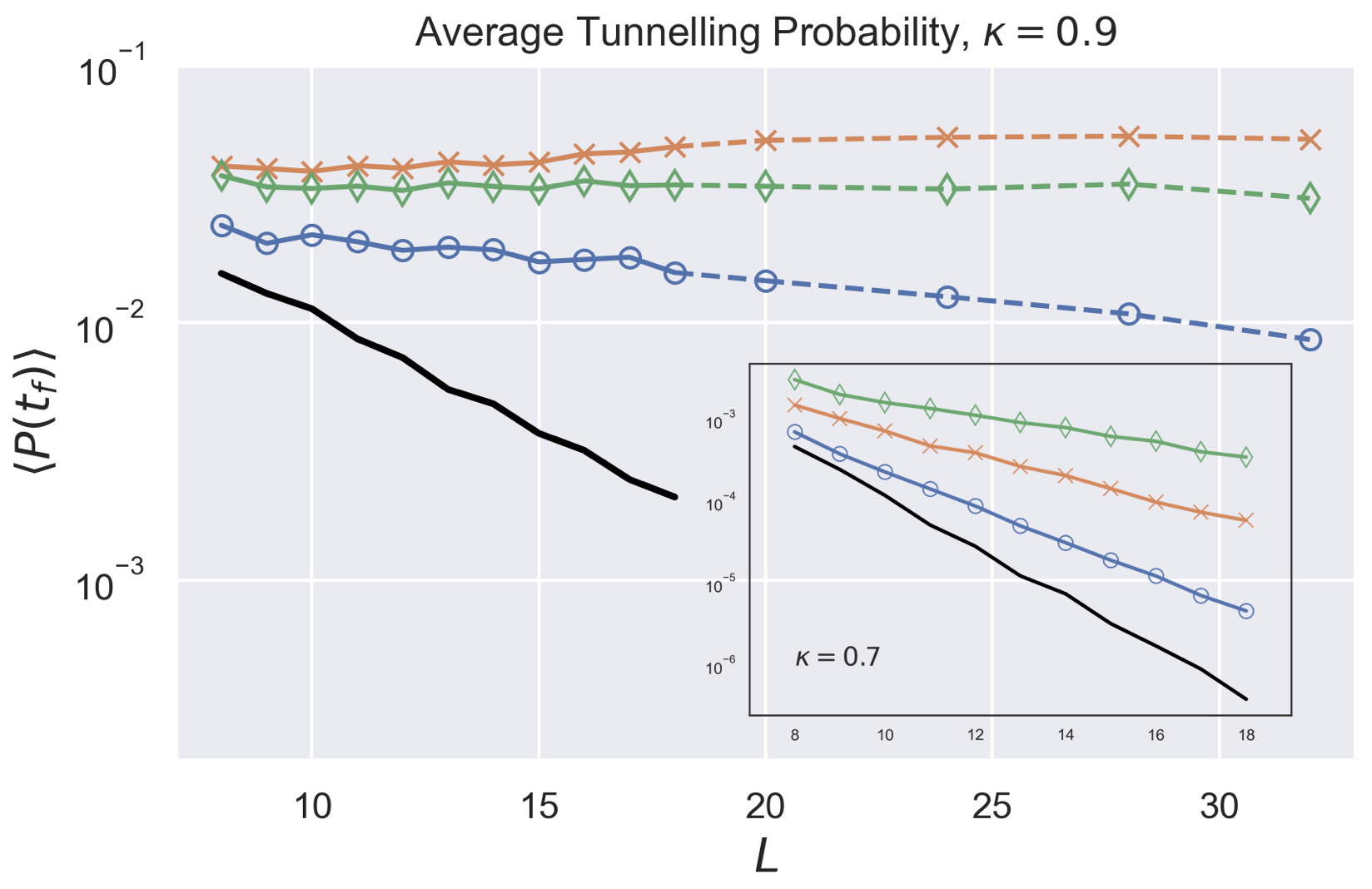}	
 
	\caption{Average tunnelling probability $\langle P(t_f)\rangle$ vs system size $L$ at $\kappa=0.9$. In the main plot: average tunnelling probability curves for the DC protocol ($\alpha=0$) and RFQA protocols with values of $\alpha$ closest to where the the difficulty exponent $\Upsilon$ is changing sign. For X- (blue dots), ZZ- (orange crosses) and XZZ-RFQA (green diamonds) these are respectively $\alpha=0.5, 0.4$ and $0.15$). Note that the average tunnelling probability for the RFQA protocols shown here is approximately constant in (or very weakly dependent on) the system size $L$, while for the DC protocol it is clearly exponentially decaying. The data for the larger system sizes $L \geq 20$ (dashed segments) were obtained using TEBD. Inset: $\langle P(t_f)\rangle$ vs $L$ plot in the positive $\Upsilon$ regime. Shown here is the case with $\kappa=0.7$, with a choice of $\alpha=0.4$ for the X- and ZZ-RFQA protocols, and $\alpha=0.2$ for the XZZ-RFQA protocol. In this regime all probability curves decay exponentially with $L$, albeit with different rates which favour the RFQA protocols over the DC-driven one.}\label{fig:tunn_accel_at_size}
	
\end{figure}

We study the transverse-field values of $\kappa=0.7,0.8,0.9$, with AC modulation amplitudes $\alpha$ between 0 and 0.5 for X-RFQA and ZZ-RFQA, and 0.3 for XZZ-RFQA. These peak amplitudes are chosen based on the observed response of the system's $L$-scaling to the various protocol choices. We observe that in all the cases we considered, RFQA oscillations significantly accelerate the global-spin-flip tunneling process at fixed scale $L$, as confirmed by the increase of the average tunnelling probability compared to the analogous DC protocol shown in Fig. \ref{fig:tunn_accel_at_size} for $L=18$, and by the exponential fits shown in Fig.~\ref{fig:tts_scaling_exp} for a wide range of parameters.

In order to capture the $L$-dependence of this improvement we study the approximate TTS ratio $t_f/\langle P(t_f) \rangle$ vs $L$ for a fixed choice of $\kappa,\alpha$ and RFQA variant (see \ref{fig:upsilon_from_TTS}). For all the DC protocols and at least the smaller-$\kappa$, smaller-$\alpha$ RFQA protocols we observe what appears \emph{prima facie} to be an exponential scaling of the TTS with $L$. We fit the approximate TTS ratio $t_f/\langle P(t_f) \rangle$ obtained from the numerical data with a two-parameter fitting function $f(L) = aL 2^{\Upsilon L}$. At small $L$ there is of course some inherent ambiguity in the choice of polynomial prefactor in such fits. We chose this form because in fits where the polynomial prefactor was allowed to vary (e.g. $L^c$ instead of $L$) the best fit exponent $c$ was always close to 1, and because the empirical scaling form in Eq.~\ref{Om0} combined with $W \propto L^{-1/2}$ predicts prefactor exponents close to 1 as well. We chose to fit all data using the same functional form for consistency and to make the easiest comparisons. From this fit we obtain the difficulty exponent $\Upsilon$. For the DC protocol, we used standard exact diagonalization methods to check that the observed TTS scaling exponent closely tracks the prediction $TTS \propto W/\Delta_{\min}^2$ obtained from the empirical gap $\Delta_{\min}$ between the two dressed quasi-degenerate ground states of the TFIM in zero longitudinal field, at fixed $\kappa$  (see Appendix \ref{appendix:min_gap_and_tts} for a demonstration of this).

We observe that for all RFQA protocols, the difficulty exponent $\Upsilon$ decreases as $\alpha$ is increased from zero to positive values (Fig. \ref{fig:tts_scaling_exp}), with XZZ-RFQA showing the most dramatic improvements and X-RFQA the most modest. For large enough values of $\kappa$ and $\alpha$, the X- and XZZ-RFQA protocols exhibit a crossover from positive to negative values of the exponent, suggesting the end of the exponentially-diverging regime of the TTS. We expect the X-RFQA variant, being the least performant of the three, will exhibit an analogous crossover at larger $\kappa,\alpha$ values of the ones we studied here. For all the three RFQA protocols we employes Time-Evolving Block Decimation (TEBD)  \cite{PhysRevLett.93.040502} methods in order to extend the calculations of the average tunnelling probability to larger system sizes $L \geq 20$, for $\kappa=0.9$ and $\alpha$ values close to the crossover point, obtaining results largely consistent with the ones at smaller $L$ (see Fig.~\ref{fig:tunn_accel_at_size}). The shifts in the fitted difficulty exponents produced by aggregating the additional TEBD data are very small: $0.048 \rightarrow 0.062$ for the X- , $-0.026 \rightarrow -0.018$ for the ZZ-, and $-0.0009 \rightarrow -0.00005$ for the XZZ-RFQA variants respectively.

\begin{table}
\centering
\begin{tabular}{| c || c | c || c | c || c | c | }
\hline
    Protocol & $\kappa$ & $\alpha$ & $\Upsilon_{ST}$ & $2 \gamma_{dis}$ & $ \avg{Z_0 Z_{L/2}}_0 $  & $ \avg{Z_0 Z_{L/2}}_{dis} $ \\
\hline
X & 0.8 & 0.5 & 0.28 & 0.39 & 0.88 & 0.87 \\
  & 0.9 & 0.5 & 0.05 & 0.15 &  0.81 & 0.8 \\
\hline
ZZ & 0.8 & 0.4 & 0.14 & 0.19 & 0.88 & 0.84 \\
  & 0.8 & 0.5 & 0.01 & 0.12 & 0.88 & 0.81 \\
  & 0.9 & 0.3 & 0.05 & 0.18 & 0.81 & 0.78 \\
  & 0.9 & 0.4 & -0.03 & 0.08 & 0.81 & 0.75 \\
\hline
XZZ & 0.8 & 0.2 & 0.05 & 0.07 & 0.88 & 0.84 \\
 & 0.8 & 0.25 & -0.03 & 0.06 & 0.88 & 0.81 \\
 & 0.9 & 0.1 & 0.11 & 0.19 & 0.81 & 0.79 \\
 & 0.9 & 0.15 & 0.0 & 0.06 & 0.81 & 0.77 \\
 \hline
\end{tabular}
\caption{Evidence for an irreducibly AC nature of tunneling acceleration in this system, by comparison of the fitted TTS exponent to the exponential decay of the DC average $\avg{\Omega_0^2}$ over the instantaneous modulation of Hamiltonian parameters appropriate to that protocol. Two point correlations are reported for $L=20$; see text for more details.}\label{DCtab}
\end{table}

\subsection{DC parameter fluctuations and ferromagnet-paramagnet phase transitions do not explain the observed speedup}

In all three of our simulated protocols, the transverse field and coupling terms are modulated fairly substantially. Consequently, in rare events, one will find appreciable shifts in the mean values of $\kappa$ or $J$, and in those cases the instantaneous collective tunnel splitting $\Omega_0$ will be much larger than in the corresponding uniform field case, potentially even large enough to induce a transition to the paramagnetic phase. A skeptical reader could very reasonably ask if rare large fluctuations are sufficient to explain the speedups we observe, and since the challenge of implementing simultaneous AC modulation of all terms is significant, it is important to rule out more prosaic explanations for the fast tunneling we report.

We first consider DC fluctuations in $\Omega_0$. As argued above, in the DC case the detuning-averaged tunneling rate $\Gamma \propto \Omega_0^2 /W$; this remains true with disorder if we replace $\Omega_0^2 \to \avg{\Omega_0^2}_{dis}$. Here, the disorder average is over random modulations of the appropriate terms equivalent to taking instantaneous time slices of $H \of{t}$ in Eq.~\ref{eq:Hamiltonian} for the appropriate protocol. To check whether this effect is able to explain the speedup, we used exact diagonalization with $h=0$ to compute $\avg{\Omega_0^2}$ for eight protocol/parameter choices (all at or close to the potential scaling crossover regime), with 2000 random disorder realizations for each datapoint and $L$ running from 8 to 20. To compare to the AC-driven tunneling rate $\Upsilon$, we numerically fit
\begin{eqnarray}\label{gapfit}
\avg{\Omega_0^2 \of{L}}  = A \frac{2^{- 2 \gamma_{dis} L}}{L}.
\end{eqnarray}
This is the same scaling form used to extract $\Upsilon$; as shown in Table~\ref{DCtab}, the resulting exponents are all larger by a shift of 0.05-0.13, well outside any fitting or sampling uncertainty here. DC fluctuations in $\Omega_0$ thus cannot explain the tunneling rates we observe. Further, since $\Omega_0$ is exponentially sensitive to mean value shifts, $\avg{\Omega_0^2}$ returns significantly larger shifts (compared to the uniform field) than $\avg{\Omega_0}$, and the \emph{median} value of $\Omega_0$ at each $L$ shows very little change compared to the uniform case.

\begin{figure}
	\centering

	\includegraphics[width=0.85\textwidth]{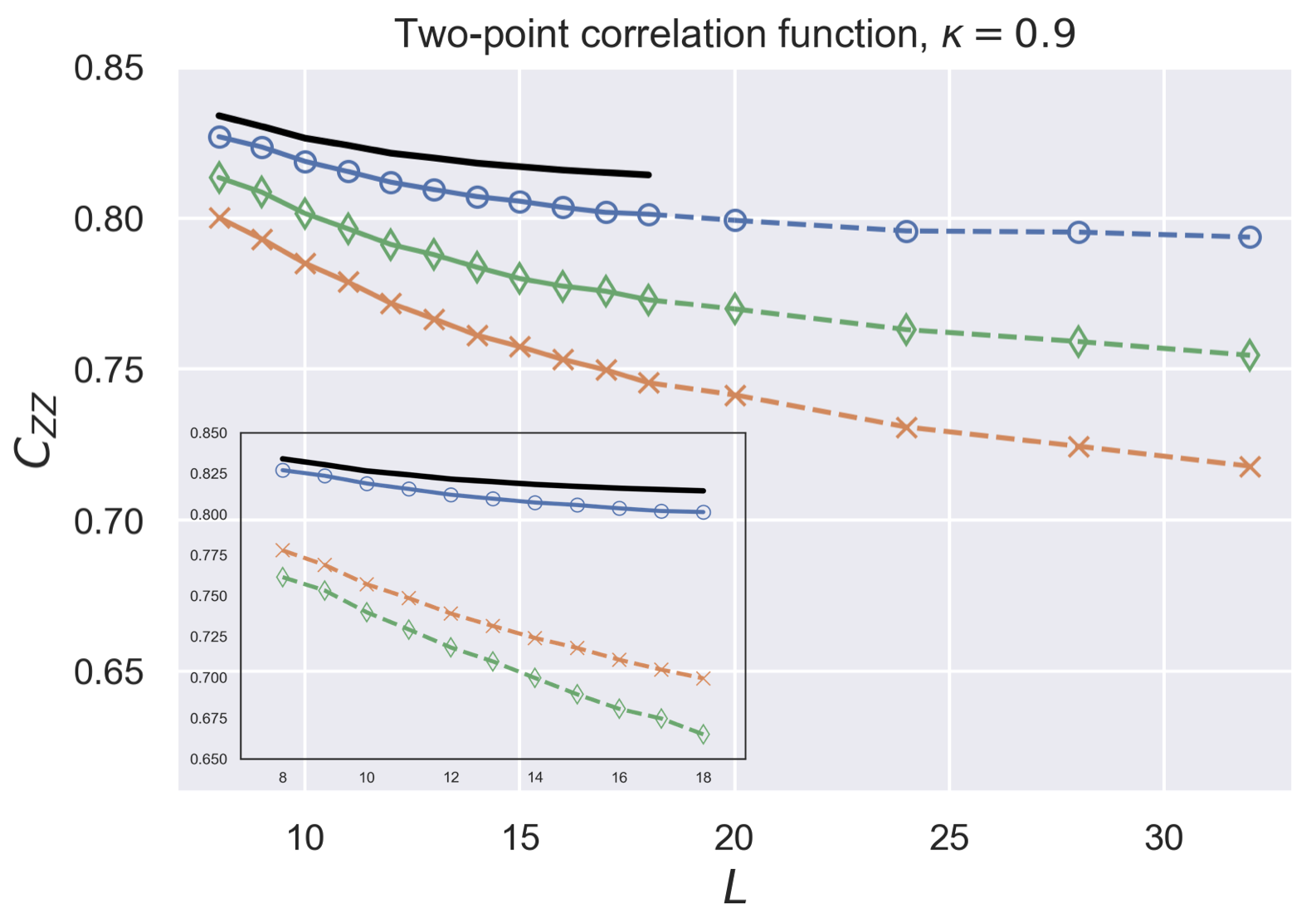}	
	\caption{Time-averaged two-point correlation function $C_{ZZ}$ in Eq.~\ref{eq:correlation_function} vs system size $L$, at fixed $\kappa=0.9$. The time-averaging is taken over the ``plateau'' part of the protocol in between the two ramps, when the global transverse field value $\kappa$ does not depend on time. The the main plot shows the curves for the DC protocol ($\alpha=0$, solid black line) and for each of the RFQA protocols, for the value of $\alpha$ closest to the where the TTS scaling exponent crosses from positive to negative values. For X- (blue dots), ZZ- (orange crosses) and XZZ-RFQA (green diamonds) these values are respectively $\alpha = 0.5$, $0.4$ and $0.15$. The data for the larger system sizes $L \geq 20$ (dashed segments) were obtained using TEBD. In the inset, the curves for the largest values of $\alpha$ we studied. For X-, ZZ- and XZZ-RFQA respectively these are $\alpha=0.5,0.5$ and $0.3$.}\label{fig:two_point_correlation}
	
\end{figure}

We now address the second possibility, transitions into the paramagnetic phase. In the clean limit this transition has a gap that scales as $1/L$; with disorder this becomes a stretched exponential \cite{fisher1995critical,dziarmaga2006dynamics,caneva2007adiabatic}, which is still substantially larger than the exponentially decaying gaps of transitions in the ferromagnetic phase. Fast tunneling driven by such transitions would not suggest utility for embedded problems, since in the paramagnetic phase long-ranged correlation (and thus, energetic awareness of the logical problem) is lost. Such transitions would also likely lead to signficant heating through the Kibble-Zurek mechanism. To ensure that this is not happening, as mentioned earlier we tracked the ferromagnetic order parameter $\avg{Z_0 Z_{L/2}}$, averaged over the ``waiting" phase of AC-driven evolution (and over detuning and frequency distributions). As shown in FIG.~\ref{fig:two_point_correlation}, and for DC eigenstates in Table~\ref{DCtab}, this order parameter is modestly reduced by the modulation terms but we see no evidence that it will vanish at large $L$. Interestingly, the AC averages are slightly \emph{larger} than the DC averages; this is because the DC average is taken at degeneracy but the AC average includes small detuning terms $h$ that bias the system further toward ferromagnetism. We thus conclude that global mixing with paramagnetic states does not explain the speedup we report.

\section{Implementation Prospects and Discussion}

Through extensive numerical simulations, we have demonstrated significant acceleration of multiqubit tunneling in transverse field Ising chains through variations of RFQA, assuming $O \of{1/\sqrt{L}}$ energy uncertainty (in comparison to an exponentially decaying minimum gap). In many of our simulations, exponential and polynomial scaling of the average tunneling time could not be distinguished from the data, in contrast to the clear exponential decay of all uniform field cases. While we reported only 1d simulations here, preliminary full state evolution studies of weakly coupled rings showed similar speedups when compared to single rings with the same total number of qubits; we will present those results in future work. As ST is not a 1d-specific mechanism, this is not surprising, and suggests these results should generalize to embedded \emph{problems}, which are far out of reach of simulation.

Of course, the ultimate goal of these AC methods is to address two of the core physics problems in quantum annealing (embedding overhead and computational weakness of a uniform transverse field), in a simple and scalable manner. While any RFQA implementation is undoubtedly more complex than current uniform transverse field hardware, as we are only varying flux-tunable $X_j / Z_j Z_k$ terms already present in the system, no direct changes to the qubit or coupler hardware are required and the complexity lies entirely on the control side. Further, we have shown that reusing the same frequencies across the lattice has performance benefits, and $O \of{\sqrt{N}}$ unique tones reused across $O \of{N}$ qubits is sufficient to rapidly accelerate tunneling while avoiding local transitions out of the ferromagnetic phase for individual chains. This favorable condition reduces our methods' implementation complexity to a local addressing and/or signal routing problem, which while difficult is still substantially easier than the expected requirements for topological error correction codes \cite{fowlersurface}. 

With regard to real implementations, it is likewise important to note that all of these simulations have been noise-free (though they do incorporate random detuning as a proxy for the many energy uncertainty sources in real experiments). This may strike the reader as a significant oversight in a paper making claims about improving near-term hardware. We chose to forgo noise simulation for a few reasons. First, we expect our results to apply (with suitable tuning and modifications) to both quantum annealers and neutral atoms, which have extremely different noise models that are both difficult to simulate in their own ways.

In neutral atoms, the error model (see the supplemental information of \cite{ebadi2022quantum}) is dominated by laser noise (amplitude and phase fluctuations of the drive beams), decay from the Rydberg state (a leakage error) and atom loss. Amplitude fluctuations\footnote{Here, we assume these fluctuations can be local or global; in previous experiments global beams were used, but obviously local control would be involved in any implementation of the protocols we discuss in this work.} in the drive lasers--weak, random modulation of the transverse fields--are going to be negligible compared to the larger modulations used to implement RFQA; phase variations correspond to slowly fluctuating $Z$ biases that stymie attempts to hit exponentially narrow resonances (at sufficiently large system size) but are less deleterious to transitions active over wider detuning ranges. Leakage and atom loss are more serious, in that they change the problem graph, and they are a common challenge to all neutral atom protocols. Of course, an RFQA implementation cannot solve these problems, though we do not expect it would meaningfully increase their rates, and since they are comparatively slow error sources, \emph{any} methods capable of finding the solution in reduced quantum evolution time can reduce their impact.

For quantum annealers, the picture is more complex, and one which was reviewed in \cite{kapit2021noise}. The error model in quantum annealers primarily consists of quasistatic control error modeled by small random fluctuations in the problem Hamiltonian parameters and transverse field strengths, $1/f$-like noise along $Z$ for each qubit, and comparatively strong interaction with a cold (but not zero temperature, e.g. $k_B T \ll J$ but $k_B T \gg \Omega_0$ at small avoided crossings) bath. RFQA does not \emph{directly} mitigate control error or $1/f$ longitudinal noise, the effects of which are at least partially captured by our detuning averaging. It is possible that RFQA could \emph{indirectly} reduce the impact of both error sources by allowing the user to work with larger intra-chain ferromagnetic coupling $J$ (which can lead to rapid freezing in the uniform field case), increasing the local gap and overall energy scale of the problem. Interaction with the cold bath is more complex, and likely prohibitively difficult to simulate in the AC-driven regime \cite{jaschke2019thermalization,kapit2020entanglement}; in a work by two of us \cite{kapit2022small}, a Hilbert space of five hundred states was used to simulate the bath interaction with two qubits, and this method does not scale to long evolution times. The original RFQA work established that AC driving accelerates bath-assisted phase transitions as well, in line with the benefits seen in uniform field annealing with a pause in the schedule \cite{marshall2019power}, but it is possible that the AC driving could also amplify harmful bath effects though an as-yet unknown mechanism. Since we cannot rigorously simulate the most important and interesting open system effects in a flux qubit implementation, we found it unnecessary to simulate other forms of noise for this work.

We want to close with an interesting question that this work raises, but certainly does not settle. Namely, the generic expectation (with some notable counterexamples \cite{laumannmoessner2012}) of first order quantum phase transitions between ground states is that the collective spin rearrangements defining the transition are exponentially slow in the number of participating degrees of freedom, e.g. $\Omega_0 \propto 2^{-c K}$. What we ask is the following: are there physically realistic--and hopefully, application relevant--first order phase transitions where, under the influence of AC driving, the nature of the transition does not change (e.g. all order parameters that are finite across the transition remain so, and the system does not meaningfully heat), but the collective tunneling rate crosses over to polynomial scaling even for arbitrarily large system sizes? Our results suggest that this is \emph{possible} but we make no claims based on numerical evidence alone that such scaling must persist at $L \to \infty$. A conclusive answer to this question would be of significant interest for quantum optimization, and dynamical many-body physics more generally.

\section{Acknowledgements}

We would like to thank Steven Dissler, Andrew King, Glen Mbeng, Eleanor Rieffel, Paul Varosy and Steven Weber for useful discussions around this project. We also would like to thank Zhijie Tang for assisting in early simulations of RFQA in 1d chains. EK and VO were jointly supported by DARPA under the Reversible Quantum Machine Learning and Simulations program, contract HR00112190068. EK's research in this area was also funded by NSF grant PHY-1653820. GM would like to acknowledge support from the NASA Ames Research Center and from DARPA under IAA 8839, Annex 128. Resources supporting this work were also provided by the NASA High-End Computing (HEC) Program through the NASA Advanced Supercomputing (NAS) Division at Ames Research Center.
\appendix


\section{Linear regime}\label{sec:linear_regime}

\begin{figure}
	\centering
	
	\includegraphics[width=0.85\textwidth]{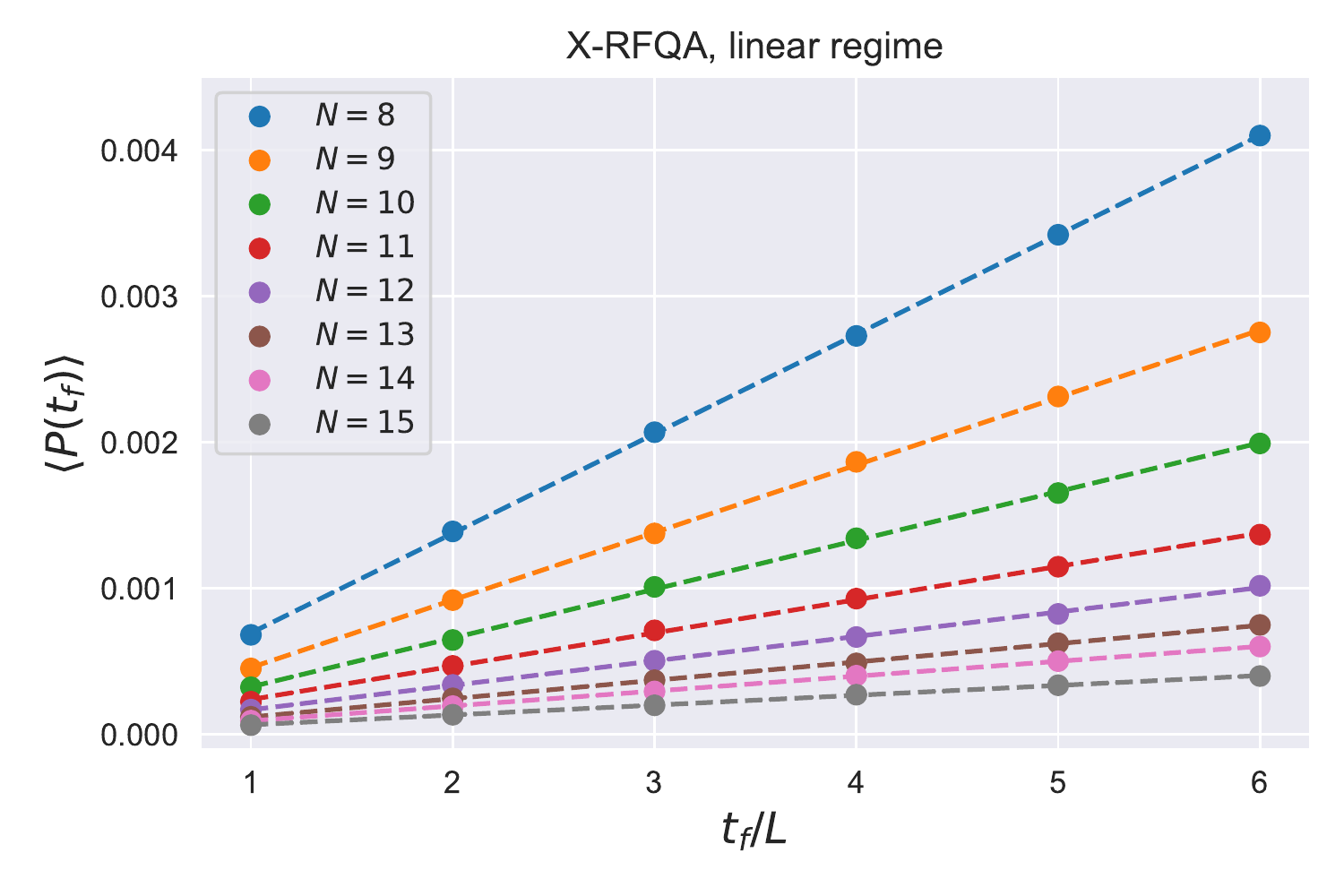}
	
	\caption{Emergence of the linear regime of the average tunnelling probability $\langle P(t_f)\rangle = \Gamma \,t_f$, for the X-RFQA protocol at $\kappa=0.8$ and $\alpha=0.3$. Dashed lines represent linear fits. The difficulty exponent $\Upsilon$ and the average rate $\Gamma$ (\emph{i.e.} the slope of the linear growth) are related through Eq.~\eqref{eq:gamma_and_upsilon_relation}.}\label{fig:linear_regime}
	
\end{figure}

The main text includes an explanation as to why one should see a linear growth regime $\langle P\of{t_f} \rangle = \Gamma \, t_f $ for the average tunnelling probability as a function of the final time of the protocols. By analogy,  the prefactor $\Gamma$ can informally be interpreted as an ``average tunnelling rate''. This linear-growth prediction is confirmed in Fig.~\ref{fig:linear_regime} by means of example, where we compare $\langle P\of{t_f} \rangle$ vs the rescaled final time $t_f/L$ in the case of X-RFQA for $\kappa=0.8,\alpha=0.3$. By using this linear-growth expression for $\langle P\of{t_f} \rangle$ in Eq.~\eqref{defTTS} one can see that in the large-$L$ limit we have that $TTS \sim 1/\Gamma$ and it would in principle be possible to extract the difficulty exponent $\Upsilon$ by studying the average rate $\Gamma$ of the linear regime. In practice, however, this becomes quickly very expensive since $\Gamma$ is typically exponentially decaying with $L$, and extracting its value requires fitting a linear function with a slope that even at moderately large values of $L$ becomes almost indistinguishable from zero. In order to do so reliably, then one would have to perform simulations for times $t_f$ that grow exponentially with $L$. For this reason we decide to extract the difficulty exponent $\Upsilon$ by fitting the scaling of the TTS directly.
 
\section{DC protocol: minimum-gap scaling exponent and difficulty exponent}\label{appendix:min_gap_and_tts}
\begin{figure}
	\centering
	
	\includegraphics[width=0.85\textwidth]{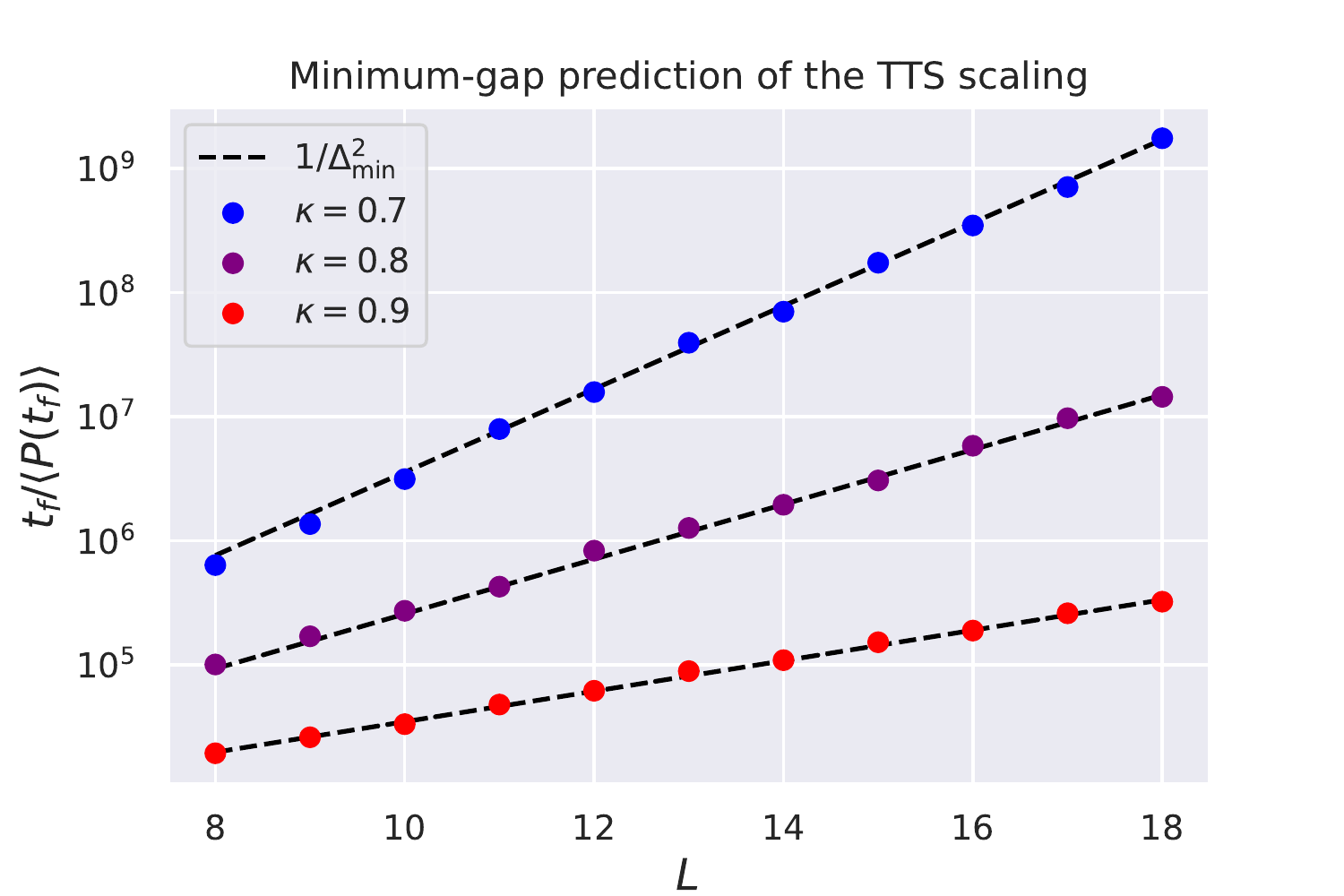}
	
	\caption{Comparison between the scaling of the TTS for the DC protocol and the inverse-square-gap prediction obtained from the ($\kappa$-dressed) empirical gap $\Delta_{min}$ between the two ferromagnetic states of the ($h=0$) transverse-field Ising chain. Datapoints correspond to dynamical data from the numerics, while dashed lines are the single-parameter fit $f(L) = b 2^{\hat{\Upsilon}L}$, where $\hat{\Upsilon}$ is the scaling exponent of the quantity $1/\Delta_{min}^2$, as explained in the main text. 
}\label{fig:mingap_squared_tts}
	
\end{figure}

We studied the scaling of the empirical gap between the two quasi-degenerate dressed ferromagnetic states at given $\kappa$ using exact diagonalization methods. We extracted the scaling exponent $\hat{\Upsilon}$ of the quantity $1/\Delta_{\min}^2$ by fitting it with the function $f(L) = a 2^{\hat{\Upsilon} L}$, with fit parameters $a,\hat{\Upsilon}$. The TTS datapoints obtained from the dynamical simulations of the DC protocol for the same given $\kappa$ were then fitted with the single-parameter function $TTS(L) = b\, 2^{\hat{\Upsilon}L}$ by using the $\hat{\Upsilon}$ value extracted before. Fig.~\ref{fig:mingap_squared_tts} shows that the scaling exponent $\hat{\Upsilon}$ of the inverse square gap $1/\Delta_{\min}^2$ corresponds to the difficulty exponent $\Upsilon$ defined by the TTS's exponential scaling with $L$. Formally,  $\lim_{L\rightarrow \infty}\frac{1}{L}\log_2 (1/\Delta^2) = \lim_{L\rightarrow \infty} \frac{1}{L} \log_2 TTS = \Upsilon$.

\section{TTS and difficulty exponent}

\begin{figure}
	\centering
	
	\includegraphics[width=1\textwidth]{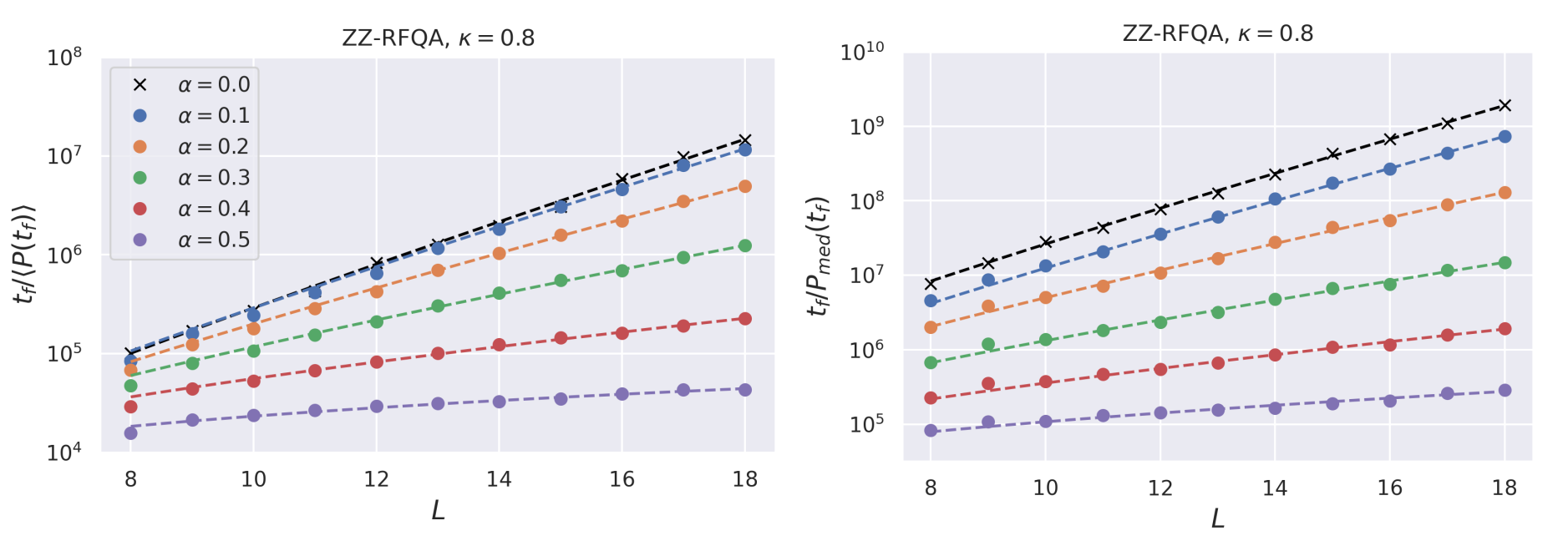}

	\caption{Left: The $L$-dependence of the approximate TTS ratio $t_f/\langle P\of{t_f}\rangle$ obtained from the numerical simuations is fitted with the two-parameter function $f(L) = aN2^{\Upsilon L}$ (dashed lines) in order to extract the difficulty exponent $\Upsilon$, in the exemplar case of the ZZ-RFQA protocol for $\kappa=0.8$ and $\alpha$ values between 0 and 0.5. Right: the ratio $t_f/P_{med}(t_f)$ defined through the \emph{median} tunnelling probability $P_{med}(t_f)$ is fitted with the same fitting function, yielding $\Upsilon_{med}$.}\label{fig:upsilon_from_TTS}
	
\end{figure}

As explained in the main text, in order to calculate the difficulty exponent $\Upsilon$ (shown in Fig.~\ref{fig:tts_scaling_exp}) for a particular protocol and fixed values of $\kappa,\alpha$,  we plot the large-$L$ approximation to the TTS, $t_f/\langle P\of{t_f}\rangle$ vs the system size $L$ and fit the data with the two-parameter fitting function $f(L) = aL 2^{\Upsilon L}$, obtaining $\Upsilon$. Fig.~\ref{fig:upsilon_from_TTS} (left) shows this for the ZZ-RFQA protocol at $\kappa=0.8$ and $\alpha$ between zero (the DC protocol) and $0.5$. The fitted $\Upsilon$ decreases monotonically with $\alpha$. Analogous behaviours are observed for all the other RFQA protocols in all parameter ranges we have studied.

The distribution of the tunneling probabilities $P(t_f)$ depends on the random choices longitudinal fields $h$, and (for the AC-driven protocols) on the random frequencies and phases. It is undesirable that this distributions -- and as a consequence, the value of the TTS -- should be dominated by rare events. If this were the case, then any observed acceleration of tunneling could be indicative of an atypical behaviour of the RFQA protocols. In order to rule out this possibility we computed the median of the tunnelling probability $P_{med}\of{t_f}$ and fitted the ratio $t_f/P_{med}\of{t_f}$ vs $L$ using the same fitting function as before. The results for the ZZ-RFQA protocol at $\kappa=0.8$ and $\alpha=0-0.5$ are shown in Fig.~\ref{fig:upsilon_from_TTS} (right). Even though we obtain slightly different scaling exponents, the overall picture is unchanged: the ``difficulty exponent'' $\Upsilon_{med}$ obtained in this way connects to the DC value at $\alpha=0$ and crosses into negative values at for large enough $\alpha$. 

\section{Heating}

\begin{figure}
	\centering
	
	\includegraphics[width=1\textwidth]{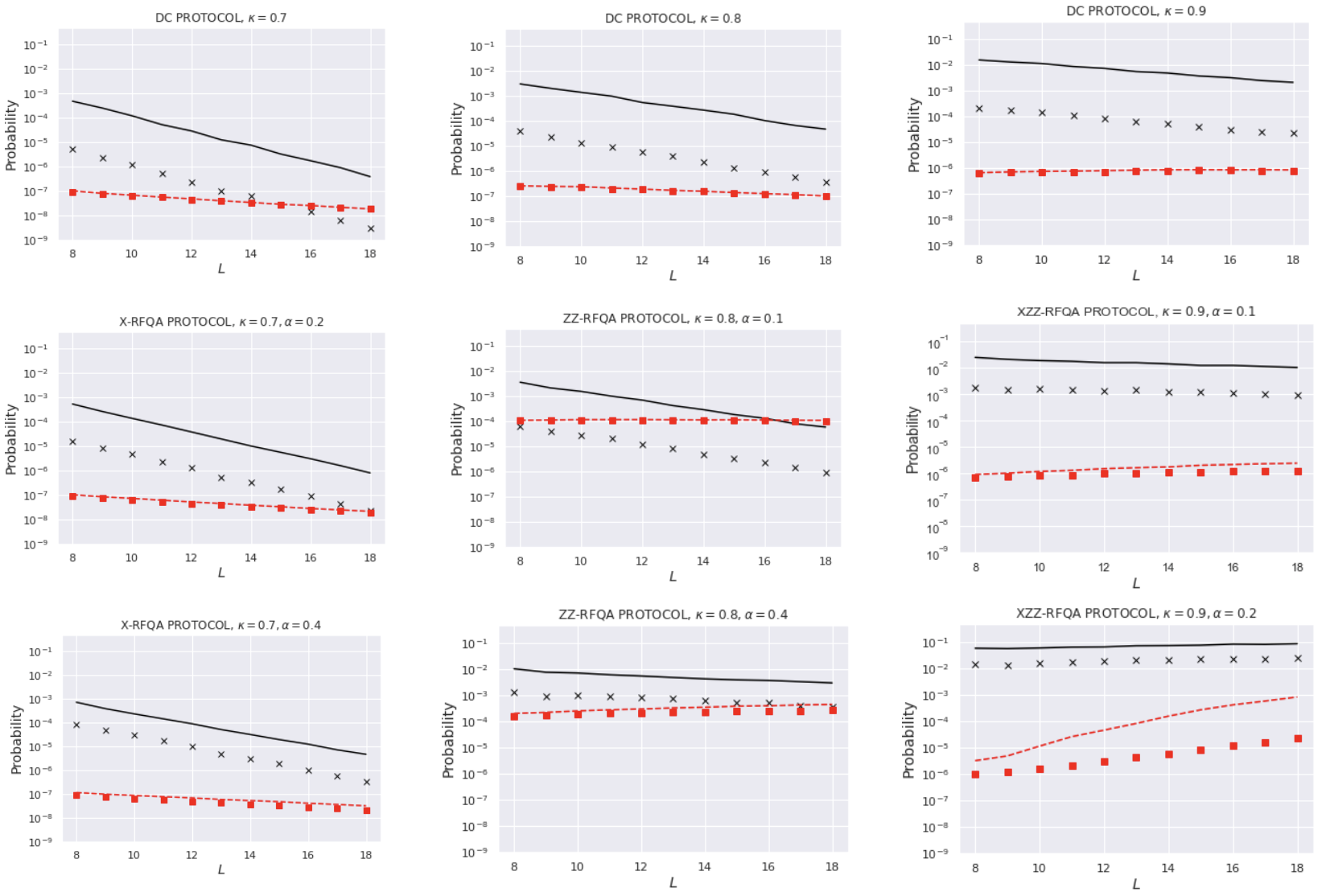}
	
	\caption{Comparisons between average $P(t_f)$ (solid black line), median $P(t_f)$ (black crosses), average $P_{heat}$ (dashed red line) and median $P_{heat}$ (red squares) in DC- and AC-driven protocols, as functions of the system size $L$. Leftmost column: X-RFQA protocol for $\kappa=0.7$ and $\alpha=0,0.2,0.4$. Center column: ZZ-RFQA protocol for $\kappa=0.8$ and $\alpha=0,0.1,0.4$. Rightmost column: ZZX-RFQA protocol for $\kappa=0.9$ and $\alpha=0,0.1,0.2$.}\label{fig:heating}
\end{figure}

In this work we use RFQA to accelerate tunnelling from one ferromagnetic state of a quantum Ising chain to the other. Ideally, this should happen without exciting the system out of the the quasi-degenerate ferromagnetic doublet. We use the \emph{heating} quantity $P_{heat}$ defined through Eq.~\eqref{eq:heating} in order to assess the amount of excitations produced by the various protocols. 
In particular, we compare $P_{heat}$ with the tunnelling probability $P\of{t_f}$ in order to rule out the possibility that $P_{heat} \gg P\of{t_f}$, which was observed in \cite{kapit2021noise} to obscure the possible advantage achieved by RFQA. For all DC protocols, and all RFQA protocols at large enough $\alpha$, we have that $\langle P_{heat} \rangle < \langle P(t_f) \rangle$ for all the system sizes studied. Some RFQA protocols go through an intermediate regime at small $\alpha$ where $\langle P\of{t_f} \rangle \lesssim \langle P_{heat} \rangle$, for the larger values of $L$. This is due to a comparatively larger amount of heat than what is observed in the DC case for the same $\kappa$, combined with too modest an increase of the tunnelling probability (See Fig.~\ref{fig:heating}). A qualitatively analogous behaviour is observed by considering the median heat and the median tunnelling probability.

Given the empirically-observed behaviour of $P(t_f)$ and $P_{heat}$, one could reasonably imagine that $P_{heat} \gg P\of{t_f}$ would eventually hold in the $L\rightarrow \infty$ limit. For the DC case, this issue can arguably be solved by using better adiabatic ramps, but for the AC-case this prediction seems hard to confirm or deny conclusively without a theoretical description of the AC-induced excitation processes that holds for non-perturbative values of $\kappa$ and $\alpha$, which we currently lack. Nevertheless, we do not expect the heating we observe \emph{at given size} to significantly affect the results presented in this work.

\section{Considerations for TEBD in this system}
For the one-dimensional model studied in this work we use Time-Evolving Block Decimation \cite{PhysRevLett.93.040502}, as implemented by the ITensor library \cite{10.21468/SciPostPhysCodeb.4,10.21468/SciPostPhysCodeb.4-r0.3} to access larger system sizes. Unlike the exact-state simulation code used for the smaller $L$, TEBD uses matrix-product states (MPS) with finite bond dimension $\chi$ in order to approximate the time-evolving wavefunction. Physical results are accessed by increasing $\chi$ until convergence is reached (up to the desired numerical accuracy). Unlike other MPS-based methods such as the Density Matrix Renormalization Group (DMRG), where the one-dimensional area law allows for an efficient representation of the ground state of a local Hamiltonian by MPSs and \emph{e.g.} for the calculation of intensive quantities (of order $O(1)$), the calculation of quantities that exhibit an exponential decay with the system size (such as the tunnelling probability $P(t_f)$ of our model) require increasingly better approximations of the time-evolving wavefunction, even in cases where the entanglement is not diverging with $L$. 

\bibliographystyle{unsrt}
\bibliography{fullbib}

\begin{thebibliography}{10}

\bibitem{finnila1994quantum}
AB~Finnila, MA~Gomez, C~Sebenik, C~Stenson, and JD~Doll.
\newblock Quantum annealing: a new method for minimizing multidimensional
  functions.
\newblock {\em Chemical physics letters}, 219(5-6):343--348, 1994.

\bibitem{kadowakinishimori1998}
Tadashi Kadowaki and Hidetoshi Nishimori.
\newblock Quantum annealing in the transverse ising model.
\newblock {\em Physical Review E}, 58(5):5355, 1998.

\bibitem{farhigoldstone2000}
E.~Farhi, J.~Goldstone, S.~Gutmann, and M.~Sipser.
\newblock Quantum computation by adiabatic evolution.
\newblock {\em arXiv:quant-ph/0001106}, 2000.

\bibitem{farhi2002quantum}
Edward Farhi, Jeffrey Goldstone, and Sam Gutmann.
\newblock Quantum adiabatic evolution algorithms versus simulated annealing.
\newblock {\em arXiv preprint quant-ph/0201031}, 2002.

\bibitem{das2008colloquium}
Arnab Das and Bikas~K Chakrabarti.
\newblock Colloquium: Quantum annealing and analog quantum computation.
\newblock {\em Reviews of Modern Physics}, 80(3):1061, 2008.

\bibitem{johnson2011quantum}
Mark~W Johnson, Mohammad~HS Amin, Suzanne Gildert, Trevor Lanting, Firas Hamze,
  Neil Dickson, R~Harris, Andrew~J Berkley, Jan Johansson, Paul Bunyk, et~al.
\newblock Quantum annealing with manufactured spins.
\newblock {\em Nature}, 473(7346):194, 2011.

\bibitem{boixo2014evidence}
Sergio Boixo, Troels~F R{\o}nnow, Sergei~V Isakov, Zhihui Wang, David Wecker,
  Daniel~A Lidar, John~M Martinis, and Matthias Troyer.
\newblock Evidence for quantum annealing with more than one hundred qubits.
\newblock {\em Nature Physics}, 10(3):218, 2014.

\bibitem{farhi2014quantum}
Edward Farhi, Jeffrey Goldstone, and Sam Gutmann.
\newblock A quantum approximate optimization algorithm.
\newblock {\em arXiv preprint arXiv:1411.4028}, 2014.

\bibitem{albashlidar2017}
Tameem Albash and Daniel~A. Lidar.
\newblock Adiabatic quantum computing.
\newblock {\em arXiv:1611.04471}, 2017.

\bibitem{albash2018demonstration}
Tameem Albash and Daniel~A Lidar.
\newblock Demonstration of a scaling advantage for a quantum annealer over
  simulated annealing.
\newblock {\em Physical Review X}, 8(3):031016, 2018.

\bibitem{scholl2021quantum}
Pascal Scholl, Michael Schuler, Hannah~J Williams, Alexander~A Eberharter,
  Daniel Barredo, Kai-Niklas Schymik, Vincent Lienhard, Louis-Paul Henry,
  Thomas~C Lang, Thierry Lahaye, et~al.
\newblock Quantum simulation of 2d antiferromagnets with hundreds of rydberg
  atoms.
\newblock {\em Nature}, 595(7866):233--238, 2021.

\bibitem{scholl2022microwave}
Pascal Scholl, Hannah~J Williams, Guillaume Bornet, Florian Wallner, Daniel
  Barredo, L~Henriet, Adrien Signoles, Cl{\'e}ment Hainaut, Titus Franz,
  S~Geier, et~al.
\newblock Microwave engineering of programmable x x z hamiltonians in arrays of
  rydberg atoms.
\newblock {\em PRX Quantum}, 3(2):020303, 2022.

\bibitem{ebadi2022quantum}
Sepehr Ebadi, Alexander Keesling, Madelyn Cain, Tout~T Wang, Harry Levine,
  Dolev Bluvstein, Giulia Semeghini, Ahmed Omran, J-G Liu, Rhine Samajdar,
  et~al.
\newblock Quantum optimization of maximum independent set using rydberg atom
  arrays.
\newblock {\em Science}, 376(6598):1209--1215, 2022.

\bibitem{perdomo2011study}
Alejandro Perdomo-Ortiz, Salvador~E Venegas-Andraca, and Al{\'a}n Aspuru-Guzik.
\newblock A study of heuristic guesses for adiabatic quantum computation.
\newblock {\em Quantum Information Processing}, 10(1):33--52, 2011.

\bibitem{king2018observation}
Andrew~D King, Juan Carrasquilla, Jack Raymond, Isil Ozfidan, Evgeny Andriyash,
  Andrew Berkley, Mauricio Reis, Trevor Lanting, Richard Harris, Fabio
  Altomare, et~al.
\newblock Observation of topological phenomena in a programmable lattice of
  1,800 qubits.
\newblock {\em Nature}, 560(7719):456--460, 2018.

\bibitem{chancellor2017modernizing}
Nicholas Chancellor.
\newblock Modernizing quantum annealing using local searches.
\newblock {\em New Journal of Physics}, 19(2):023024, 2017.

\bibitem{marshall2019power}
Jeffrey Marshall, Davide Venturelli, Itay Hen, and Eleanor~G Rieffel.
\newblock Power of pausing: Advancing understanding of thermalization in
  experimental quantum annealers.
\newblock {\em Physical Review Applied}, 11(4):044083, 2019.

\bibitem{king2023quantum}
Andrew~D King, Jack Raymond, Trevor Lanting, Richard Harris, Alex Zucca, Fabio
  Altomare, Andrew~J Berkley, Kelly Boothby, Sara Ejtemaee, Colin Enderud,
  et~al.
\newblock Quantum critical dynamics in a 5,000-qubit programmable spin glass.
\newblock {\em Nature}, pages 1--6, 2023.

\bibitem{choi2008minor}
Vicky Choi.
\newblock Minor-embedding in adiabatic quantum computation: I. the parameter
  setting problem.
\newblock {\em Quantum Information Processing}, 7(5):193--209, 2008.

\bibitem{choi2011minor}
Vicky Choi.
\newblock Minor-embedding in adiabatic quantum computation: Ii. minor-universal
  graph design.
\newblock {\em Quantum Information Processing}, 10(3):343--353, 2011.

\bibitem{konz2021embedding}
Mario~S K{\"o}nz, Wolfgang Lechner, Helmut~G Katzgraber, and Matthias Troyer.
\newblock Embedding overhead scaling of optimization problems in quantum
  annealing.
\newblock {\em PRX Quantum}, 2(4):040322, 2021.

\bibitem{kim2021rydberg}
Minhyuk Kim, Kangheun Kim, Jaeyong Hwang, Eun-Gook Moon, and Jaewook Ahn.
\newblock Rydberg quantum wires for maximum independent set problems with
  nonplanar and high-degree graphs.
\newblock {\em arXiv preprint arXiv:2109.03517}, 2021.

\bibitem{nguyen2023quantum}
Minh-Thi Nguyen, Jin-Guo Liu, Jonathan Wurtz, Mikhail~D Lukin, Sheng-Tao Wang,
  and Hannes Pichler.
\newblock Quantum optimization with arbitrary connectivity using rydberg atom
  arrays.
\newblock {\em PRX Quantum}, 4(1):010316, 2023.

\bibitem{marshall2017thermalization}
Jeffrey Marshall, Eleanor~G Rieffel, and Itay Hen.
\newblock Thermalization, freeze-out, and noise: Deciphering experimental
  quantum annealers.
\newblock {\em Physical Review Applied}, 8(6):064025, 2017.

\bibitem{hamerly2019experimental}
Ryan Hamerly, Takahiro Inagaki, Peter~L McMahon, Davide Venturelli, Alireza
  Marandi, Tatsuhiro Onodera, Edwin Ng, Carsten Langrock, Kensuke Inaba,
  Toshimori Honjo, et~al.
\newblock Experimental investigation of performance differences between
  coherent ising machines and a quantum annealer.
\newblock {\em Science advances}, 5(5):eaau0823, 2019.

\bibitem{kowalsky20213}
Matthew Kowalsky, Tameem Albash, Itay Hen, and Daniel~A Lidar.
\newblock 3-regular 3-xorsat planted solutions benchmark of classical and
  quantum heuristic optimizers.
\newblock {\em arXiv preprint arXiv:2103.08464}, 2021.

\bibitem{kapit2021noise}
Eliot Kapit and Vadim Oganesyan.
\newblock Noise-tolerant quantum speedups in quantum annealing without fine
  tuning.
\newblock {\em Quantum Science and Technology}, 6(2):025013, 2021.

\bibitem{zalka1999}
Christof Zalka.
\newblock Grover's quantum searching algorithm is optimal.
\newblock {\em Physical Review A}, 60(4):2746, 1999.

\bibitem{isakov2016understanding}
Sergei~V Isakov, Guglielmo Mazzola, Vadim~N Smelyanskiy, Zhang Jiang, Sergio
  Boixo, Hartmut Neven, and Matthias Troyer.
\newblock Understanding quantum tunneling through quantum monte carlo
  simulations.
\newblock {\em Physical review letters}, 117(18):180402, 2016.

\bibitem{andriyash2017can}
Evgeny Andriyash and Mohammad~H Amin.
\newblock Can quantum monte carlo simulate quantum annealing?
\newblock {\em arXiv preprint arXiv:1703.09277}, 2017.

\bibitem{jiang2017scaling}
Zhang Jiang, Vadim~N Smelyanskiy, Sergei~V Isakov, Sergio Boixo, Guglielmo
  Mazzola, Matthias Troyer, and Hartmut Neven.
\newblock Scaling analysis and instantons for thermally assisted tunneling and
  quantum monte carlo simulations.
\newblock {\em Physical Review A}, 95(1):012322, 2017.

\bibitem{jiang2017path}
Zhang Jiang, Vadim~N Smelyanskiy, Sergio Boixo, and Hartmut Neven.
\newblock Path-integral quantum monte carlo simulation with open-boundary
  conditions.
\newblock {\em Physical Review A}, 96(4):042330, 2017.

\bibitem{king2019scaling}
Andrew~D King, Jack Raymond, Trevor Lanting, Sergei~V Isakov, Masoud Mohseni,
  Gabriel Poulin-Lamarre, Sara Ejtemaee, William Bernoudy, Isil Ozfidan,
  Anatoly~Yu Smirnov, et~al.
\newblock Scaling advantage in quantum simulation of geometrically frustrated
  magnets.
\newblock {\em arXiv preprint arXiv:1911.03446}, 2019.

\bibitem{crosson2014different}
Elizabeth Crosson, Edward Farhi, Cedric Yen-Yu Lin, Han-Hsuan Lin, and Peter
  Shor.
\newblock Different strategies for optimization using the quantum adiabatic
  algorithm.
\newblock {\em arXiv preprint arXiv:1401.7320}, 2014.

\bibitem{PhysRevLett.93.040502}
Guifr\'e Vidal.
\newblock Efficient simulation of one-dimensional quantum many-body systems.
\newblock {\em Phys. Rev. Lett.}, 93:040502, Jul 2004.

\bibitem{PhysRevLett.91.147902}
Guifr\'e Vidal.
\newblock Efficient classical simulation of slightly entangled quantum
  computations.
\newblock {\em Phys. Rev. Lett.}, 91:147902, Oct 2003.

\bibitem{PAECKEL2019167998}
Sebastian Paeckel, Thomas Köhler, Andreas Swoboda, Salvatore~R. Manmana,
  Ulrich Schollwöck, and Claudius Hubig.
\newblock Time-evolution methods for matrix-product states.
\newblock {\em Annals of Physics}, 411:167998, 2019.

\bibitem{10.21468/SciPostPhysCodeb.4}
Matthew Fishman, Steven~R. White, and E.~Miles Stoudenmire.
\newblock {The ITensor Software Library for Tensor Network Calculations}.
\newblock {\em SciPost Phys. Codebases}, page~4, 2022.

\bibitem{10.21468/SciPostPhysCodeb.4-r0.3}
Matthew Fishman, Steven~R. White, and E.~Miles Stoudenmire.
\newblock {Codebase release 0.3 for ITensor}.
\newblock {\em SciPost Phys. Codebases}, pages 4--r0.3, 2022.

\bibitem{PhysRevA.92.042325}
Itay Hen, Joshua Job, Tameem Albash, Troels~F. R\o{}nnow, Matthias Troyer, and
  Daniel~A. Lidar.
\newblock Probing for quantum speedup in spin-glass problems with planted
  solutions.
\newblock {\em Phys. Rev. A}, 92:042325, Oct 2015.

\bibitem{fisher1995critical}
Daniel~S Fisher.
\newblock Critical behavior of random transverse-field ising spin chains.
\newblock {\em Physical review b}, 51(10):6411, 1995.

\bibitem{dziarmaga2006dynamics}
Jacek Dziarmaga.
\newblock Dynamics of a quantum phase transition in the random ising model:
  Logarithmic dependence of the defect density on the transition rate.
\newblock {\em Physical Review B}, 74(6):064416, 2006.

\bibitem{caneva2007adiabatic}
Tommaso Caneva, Rosario Fazio, and Giuseppe~E Santoro.
\newblock Adiabatic quantum dynamics of a random ising chain across its quantum
  critical point.
\newblock {\em Physical Review B}, 76(14):144427, 2007.

\bibitem{fowlersurface}
A.~G. Fowler, M.~Mariantoni, J.~M. Martinis, and A.~N. Cleland.
\newblock Surface codes: Towards practical large-scale quantum computation.
\newblock {\em Phys. Rev. A \textbf{86}, 032324}, 2012.

\bibitem{jaschke2019thermalization}
Daniel Jaschke, Lincoln~D Carr, and In{\'e}s de~Vega.
\newblock Thermalization in the quantum ising model?approximations, limits, and
  beyond.
\newblock {\em Quantum Science and Technology}, 4(3):034002, 2019.

\bibitem{kapit2020entanglement}
Eliot Kapit, Pedram Roushan, Charles Neill, Sergio Boixo, and Vadim
  Smelyanskiy.
\newblock Entanglement and complexity of interacting qubits subject to
  asymmetric noise.
\newblock {\em Physical Review Research}, 2(4):043042, 2020.

\bibitem{kapit2022small}
Eliot Kapit and Vadim Oganesyan.
\newblock Small logical qubit architecture based on strong interactions and
  many-body dynamical decoupling.
\newblock {\em arXiv preprint arXiv:2212.04588}, 2022.

\bibitem{laumannmoessner2012}
C.~Laumann, R.~Moessner, A.~Scardicchio, and S.~L. Sondhi.
\newblock The quantum adiabatic algorithm and scaling of gaps at first order
  quantum phase transitions.
\newblock {\em Phys. Rev. Lett. \textbf{109}, 030502}, 2012.

\end{thebibliography}

\end{document}